\documentclass[11pt,a4paper]{article}
\usepackage{geometry}
\usepackage[ngerman,english]{babel}
\usepackage{mathrsfs}
\usepackage{bbm}
\usepackage{amsmath}
\usepackage{amssymb}
\usepackage{amsfonts}
\usepackage{jcap}
\usepackage{setspace}
\usepackage{color,soul}
\newcommand{\z}{\zeta}
\newcommand{\zs}{\zeta_{s}}
\newcommand{\zl}{\zeta_{l}}
\newcommand{\f}{f_{\text{NL}}}
\newcommand{\g}{g_{\text{NL}}}
\newcommand{\fn}{f_{\text{NL}}^{\text{N}}}
\newcommand{\ph}{\phi_{1}}
\newcommand{\ps}{\phi_{s}}

\newcommand{\HH}{\ensuremath{\mathcal{H}}}
\newcommand{\F}{f_{1}\left(\Omega_{m}\right)+\frac{3}{2}\Omega_{m}}
\newcommand{\tht}{\vartheta}

\newcommand{\RM}[2]{#1_{\rm #2}}
\newcommand{\ze}[1]{\zeta^{(#1)}}

\newcommand{\cg}{\check{\gamma}}

\newcommand{\pll}{\phi_{l}}

\title{\boldmath$f_{\text{NL}}-g_{\text{NL}}$ mixing in the matter density field at higher orders}
\author[a]{Hedda A. Gressel} 
\author[a]{Marco Bruni%
}
\affiliation[a]{Institute of Cosmology and Gravitation, University of
  Portsmouth\\
   Dennis Sciama Building, Portsmouth PO1 3FX, United Kingdom}
\emailAdd{hedda.gressel@port.ac.uk}
\emailAdd{marco.bruni@port.ac.uk}
\abstract{
In this paper we examine how primordial non-Gaussianity  contributes to nonlinear perturbative orders  in the expansion of the density field at  large scales in the matter dominated era. General Relativity is an intrinsically nonlinear theory, establishing a nonlinear relation between the metric and the density field. Representing the metric perturbations with the curvature perturbation $\zeta$, it is known  that nonlinearity produces effective non-Gaussian terms in the nonlinear perturbations of the matter density field $\delta$, even if the primordial $\zeta$ is Gaussian.  Here we generalise these results to the case of a non-Gaussian primordial $\zeta$.   Using a standard parametrization of primordial non-Gaussianity in $\zeta$ in terms of  $f_{\text{NL}}$, $g_{\text{NL}}$,  $h_{\text{NL}}$..., we show how at higher order (from third and higher) nonlinearity also produces a mixing of these contributions to the density field at large scales, e.g.\ both  $f_{\text{NL}}$ and $g_{\text{NL}}$ contribute to the third order in $\delta$. This is the main result of this paper. 
Our analysis is based on the synergy between a gradient expansion  (aka long-wavelength approximation) and standard perturbation theory at higher order. In essence, mathematically the equations for the gradient expansion are equivalent to those of first order perturbation theory, thus first-order results convert into gradient expansion results and, vice versa, the gradient expansion can be used to derive results in perturbation theory at higher order and large scales.

}
\arxivnumber{XXX}
\begin{document}
\maketitle
\flushbottom
\section{Introduction}
\label{sec:intro}
Non-Gaussianity of primordial fluctuations, a residue from the inflationary era, is a powerful probe of the dynamics of the very early universe. The bispectrum of the cosmic microwave background radiation (CMB) provides the statistical measure for non-Gaussianity and insights into the conditions in the inflationary universe \cite{Bartolo:2004if,Wands:2010af,Koyama:2010xj,Byrnes:2010em}.
Recently, high precision measurements with the Planck satellite were able to further constraint the value of the local type non-Gaussianity $f_{\text{NL}}$ \cite{Ade:2015ava}. In upcoming galaxy surveys, 
primordial non- Gaussianity will be probed thanks to  its  scale-dependence on large scales \cite{2008PhRvD..77l3514D}, where however it is important to consider relativistic effects \cite{Bruni:2011ta}.

However,  even with Gaussian primordial                       fluctuations, the intrinsic nonlinearity of General Relativity produces non-Gaussian contributions in the matter density field \cite{2041-8205-794-1-L11,Matarrese:1997ay,Bernardeau:2001qr,Bartolo:2005fp,2010CQGra..27l4009B}. In particular \cite{M14,Villa:2015ppa} show how this effective non-Gaussianity and primordial non-Gaussianity add to the evolution of the density field up to second order. There have been recent discussions on the topic whether and how this effective non-Gaussianity contributes to the galaxy  bias \cite{Bartolo:2015qva,Dai:2015jaa,dePutter:2015vga,Desjacques:2016bnm}; however, in this paper we restrict our attention to the underlying matter density field.

We use  the gradient expansion approximation scheme, also known as long-wavelength approximation \cite{0038-5670-6-4-A04,Tom,PhysRevD.31.1792,PhysRevD.42.3936,PhysRevD.52.2007,0264-9381-20-24-003,Rampf:2013ewa,2041-8205-794-1-L11},   to  investigate non-Gaussian contribution in the density field at very large scales, up to fourth order  in standard perturbation theory, in the context of standard $\Lambda$CDM cosmology.
Thus, we focus on scales large enough to neglect spatial gradients in comparison to the time derivatives. We discuss the contributions derived  from the nonlinear nature of General Relativity as well as primordial non-Gaussianity up to fourth order.
To describe collisionless  matter, CDM,  we consider a pressureless irrotational dust flow  in synchronous-comoving gauge. 
The outline of the paper is as follows. In section \ref{sec:2}, we summarise the essential exact equations that are needed in the following sections to study  the nonlinear evolution  of the density contrast. In later sections, we perturb and expand these equations, i.e.\  the exact continuity equation for the density contrast, the exact Raychaudhuri equation for the expansion scalar, and the exact energy constraint that links density contrast and expansion scalar to the spatial curvature. In general these equations are then nonlinearly coupled to the equations for the shear of the matter flow and for the Weyl tensor \cite{Ellis:2009}, but in the approximation used in this paper these three equations are all is needed, at any perturbative order and at large scales, as we are going to show.
Section \ref{sec:GE} is dedicated to the gradient expansion. We omit any quantities of order higher than $\mathcal{O}(\nabla^{2})$.  
By splitting scales into  long and short, we can safely approximate the  evolution of our observable part of the universe (see Appendix \ref{App:A}) as that of 
 a separate (homogeneous) universe with its own  background density and curvature. This is commonly regarded as the separate universe conjecture.
Within this approximation, the metric reduces to a conformally flat metric with an effective scale factor constructed from the scale factor $a$ and the metric perturbation $\zeta$. Within the approximation of the gradient expansion, the quantities still contain all orders from standard perturbation theory (SPT). 
However, in section \ref{sec:CPA} we contrast the results of section \ref{sec:GE} with SPT and study the first and second order of the evolution equations. Thereby, we find that the first-order equations of SPT coincide with the approximation of the gradient expansion.
In section \ref{sec:34O}, we formulate the density contrast in terms of a series expansion and compute all orders up to order $\mathcal{O}(4)$ in SPT. In addition, we add non-Gaussian contributions up to fourth order in the initial conditions to examine the evolution of non-Gaussian contributions reflecting on possible inflationary scenarios.
Appendix \ref{App:A} covers the spatial Ricci scalar in terms of the metric. We argue which contributions are kept and which neglected once we perform a gradient expansion up to order $\mathcal{O}\left(\nabla^2 \right)$.  In the appendix \ref{App:B} and \ref{App:C} we relate the first order curvature perturbation $\zeta^{(1)}$ to the Poisson gauge metric potentials in order to subsequently compare our results to those in the Newtonian approximation. 
\section{\boldmath Evolution equation for the density contrast \texorpdfstring{$\delta$}{ d}} 
\label{sec:2}
In this section we will provide the basis for deriving the evolution equations for the density contrast using the Einstein field equations, the deformation tensor, and the continuity equation for the density contrast.

A general cosmological line element can be written as
  \begin{equation}
  ds^{2}=a^{2}(\eta)\left[-\left(1+2\phi \right)d\eta^{2}+2\omega_{i}d\eta dx^{i}+\gamma_{ij}dx^{i}dx^{j}\right],
  \end{equation}
where $\eta$ is the conformal time, $a(\eta)$ the scale factor and $\gamma_{ij}$ is the conformal spatial metric.\\
From now on, we will use the synchronous-comoving gauge, so that $\phi=\omega_{i}=0$ \cite{Landau-Lifshitz} (see Appendices \ref{App:B} and \ref{App:C} for relations to other gauges).

We consider a pressureless, irrotational fluid and comoving observers with four-velocity $u_{\mu}=(-a,0,0,0)$. Thus, the four-velocity $u^{\mu}$ of the fluid and of the observers coincides with the normal $n^{\mu}$ of constant time hypersurfaces.  Using $u^{\mu}$ we can covariantly define kinematical quantities, following the covariant fluid approach \cite{PhysRevD.40.1804,Ellis:2009,Ellis:2012}; the projection tensor $h^{\mu}_{\, \nu}$ coincides with the spatial metric  $h^{\mu}_{\, \nu}\equiv g^{\mu}_{\, \nu}+u^{\mu}u_{\nu}$ in the constant time hypersurfaces.

The deformation tensor of the fluid is defined as 
  \begin{equation}
  \vartheta^{\mu}_{\, \nu}\equiv a u^{\mu}_{\, ;\nu}-\HH h^{\mu}_{\, \nu}, \label{eq:deften}
  \end{equation}
where $\HH=a'/a$ is the conformal Hubble scalar, the prime indicates conformal time derivative,  and the isotropic background expansion $3\HH$ has been subtracted. Then, the trace $\vartheta=\vartheta^{\mu}_{\, \mu}$ of the deformation tensor denotes the inhomogeneous volume expansion and the traceless part represents the matter shear tensor.

Due to our synchronous-comoving gauge choice, with $u^{\mu}=n^{\nu}$, the deformation tensor is purely spatial and coincides with the negative of the extrinsic curvature $K^{i}_{\, j}$ of the conformal spatial metric $\gamma_{ij}$, which can be expressed as follows \cite{Wald:1984rg}:
  \begin{equation}
  \vartheta^{i}_{\,j}=-K^{i}_{\, j}\equiv \frac{1}{2}\gamma^{ik}\gamma'_{kj}.\label{eq:thk}
  \end{equation}
 The matter density field is characterised by a background part $\bar{\rho}$ and a density contrast $\delta$, as in equation \eqref{eq:deften} for the deformation tensor, 
  \begin{equation}
  \rho(\mathbf{x},\eta)=\bar{\rho}(\eta)+\delta\rho(\mathbf{x},\eta)=\bar{\rho}(\eta)\left(1+\delta(\mathbf{x},\eta)\right). \label{eq:delta0}
  \end{equation}
Using \eqref{eq:delta0}, the energy conservation equation $u_{\alpha}T^{\alpha \beta}_{\hspace{2mm};\beta}=0$ gives the continuity equation for the density contrast:
  \begin{equation}
  \delta'+\left(1+\delta\right)\vartheta=0.\label{eq:cont}
  \end{equation}
The evolution equation for the expansion, $\tht$, is given by the Raychaudhuri equation:
\begin{equation}
  \tht'+\HH\tht+\tht^{i}_{j}\tht^{j}_{i}+4 \pi G a^{2}\bar{\rho}\delta=0.\label{eq:rayy}
  \end{equation}
Furthermore, via \eqref{eq:thk} one obtains the energy constraint  \cite{Wald:1984rg,Ellis:2012}:
  \begin{equation}
  \tht^{2}-\tht^{i}_{\, j}\tht^{j}_{\, i}+4\HH \tht+ R=16 \pi G a^{2} \bar{\rho}\delta, \label{eq:energcon}
  \end{equation}
where $R$ refers to the purely spatial Ricci scalar of the conformal spatial metric $\gamma_{ij}$. This is the 00 component of the Einstein field equations in the synchronous-comoving gauge.

Both in equation \eqref{eq:rayy} and \eqref{eq:energcon}, the term $\tht^{i}_{j}\tht^{j}_{i}$ couples these equations to the evolution equations of the shear and the Weyl tensor \cite{PhysRevD.40.1804,Ellis:2009,Ellis:2012}.
However, in the approximation used in the following, we only need the equations above.
    \section{The gradient expansion}
    \label{sec:GE}
We choose the following representation of the spatial metric:
  \begin{equation}
 g_{ij}=a^{2} \gamma_{ij}=a^{2}e^{2\zeta}\check{\gamma}_{ij}, \label{eq:metric}
  \end{equation}
where $\zeta$ denotes the primordial curvature perturbation. This variable is customarily used to to deal with primordial non-Gaussianity from inflation \cite{Wands:2010af} (see Appendix \ref{App:B} for a first-order gauge-invariant treatment). Furthermore, we only consider scalar perturbations.

In the standard model of cosmology, $\Lambda$CDM, it is assumed that inflation imposes the initial condition in the very early universe. For scalar perturbations, initial conditions are given by the primordial curvature perturbation $\zeta$. This is convenient, because $\zeta$ remains constant after inflation ends and is almost scale-invariant \cite{lyth2009primordial}.
By performing a gradient expansion up to second order, we focus on scales large enough that the spatial gradients are small compared to time derivatives and terms of order higher than $\mathcal{O}(\nabla^{2})$ are negligible, where $\nabla$ is the spatial gradient in comoving coordinates
 \cite{0038-5670-6-4-A04,Tom,PhysRevD.31.1792,PhysRevD.42.3936,PhysRevD.52.2007,0264-9381-20-24-003, Rampf:2013ewa,2041-8205-794-1-L11}. In this approximation, one finds that
  \begin{equation}
  \delta \sim \vartheta \sim R \sim \nabla^{2}.
  \end{equation}
Consequently, the continuity equation \eqref{eq:cont} and the energy constraint \eqref{eq:energcon} become
  \begin{equation}
  \delta'+\tht=\mathcal{O}(\nabla^{4}) \label{eq:cont2}
  \end{equation}
and 
  \begin{equation}
  4\HH \tht +R=16\pi G a^{2}\bar{\rho}\delta +\mathcal{O}(\nabla^{4}) \label{eq:energcon2}.
  \end{equation}
Note that we have not perturbed any quantity in the conventional sense. Thus, from the point of view of the standard perturbative approach $R$, $\delta$, and $\tht$ are nonlinear and contain all orders (at large scales).\\
We now combine \eqref{eq:cont2} and \eqref{eq:energcon2} and, thereby, formulate an evolution equation for the density contrast:
  \begin{equation}
  4\HH \delta'- R=16\pi G a^{2}\bar{\rho}\delta \label{eq:diffglei}.
  \end{equation}
$R$ remains constant, i.e.\ it is a conserved quantity  in this large-scale approximation. This can easily be seen by taking the time derivative of \eqref{eq:energcon2} and combining the result with the Raychaudhuri equation \eqref{eq:rayy}, which reads
\begin{equation}
  \tht'+\HH\tht+4 \pi G a^{2}\bar{\rho}\delta=0.\label{eq:rayylin}
  \end{equation}
  within the gradient expansion up to $\mathcal{O}\left(\nabla^2\right)$.
  
The crucial  feature of  the above equations is that they take the same form as the first order equations in the standard perturbative approach (cf.\ section \ref{sec:CPA} and \cite{M14}). Therefore, the evolution equation \eqref{eq:diffglei} and its solution are formally equivalent to the first order evolution equation and solution, respectively. In the standard perturbation framework, the first order Ricci scalar\footnote{In order to distinguish between the different approximation schemes, we refer with $R^{(i)}$ to  the order $i$ of $R$ in the standard perturbative approach.}, $R^{(1)}$, is conserved. At second order, $R^{(2)}$ comprises a time dependent and a conserved part. In the gradient expansion, we only take the conserved contribution of each order $i$ of $R^{(i)}$ into account. 

At leading order on large scales, we can safely approximate the spatial metric as $\check{\gamma}_{ij}\simeq \delta_{ij}$, because non-flat contributions to  $\check{\gamma}_{ij}$ are higher order in the gradient expansion (see appendix \ref{App:A}). Hence, in this approximation the spatial metric \eqref{eq:metric} is conformally flat, with the conformal factor $a^{2}e^{2\zeta}$; this conformal factor can be seen as an effective scale factor in the separate universe approach \cite{Lem,Dai:2015jaa}.
Given the conformal flatness of the spatial metric, the Ricci scalar is a nonlinear function solely of the curvature perturbation $\zeta$ and takes on the form \cite{Wald:1984rg,M14,2041-8205-794-1-L11}
\begin{align}
R=e^{-2\zeta}\left[-4\nabla^{2}\zeta-2\left(\nabla\zeta\right)^{2}\right]. \label{eq:expR}
\end{align}
Performing a series expansion of the exponential, \eqref{eq:expR} yields   
\begin{align}
R=& \sum^{\infty}_{n=0}\frac{(-2\zeta)^{n}}{n!}\left[-4\nabla^{2}\zeta-2\left(\nabla\zeta\right)^{2}\right]\notag\\
=&-4\nabla^{2}\zeta-\sum_{n=1}^{\infty}\frac{(-2\zeta)^{n}}{n!}4\nabla^{2}\zeta-\sum^{\infty}_{n=0}\frac{(-2\zeta)^{n}}{n!}2\left(\nabla\zeta\right)^{2}\notag\\
=&-4\nabla^{2}\zeta+\sum_{n=0}^{\infty}\frac{(-2)^{n+1}}{(n+1)!}\left[-4\zeta\nabla^{2}\zeta+(n+1)\left(\nabla\zeta\right)^{2}\right]\zeta^{n}.\label{eq:RExp}
\end{align}
As we shall see in section \ref{sec:34O} this can be used to represent $R$ up to any desired perturbative order in $\zeta$.

  \section{Contrasting with the standard perturbative approach}
\label{sec:CPA}
So far we have used a gradient expansion, keeping the leading order, rather than applying the standard perturbative expansion to quantities representing inhomogeneities. 
We now clarify the relationship between the two approaches. \cite{M14,2041-8205-794-1-L11}

In the standard perturbative expansion, combining the first-order parts of \eqref{eq:cont} and \eqref{eq:energcon},
 one obtains the following evolution equation for the density contrast:
\begin{equation}
4\HH \delta^{(1)\prime}+6\HH^{2}\Omega_{m}\delta^{(1)}- R^{(1)}=0 \label{eq:diffglei1o},
\end{equation}
with 
\begin{equation}
R^{(1)\prime}=0.\label{eq:R1}
\end{equation}
Equation \eqref{eq:diffglei1o} has exactly the same form of the evolution equation \eqref{eq:diffglei} obtained in the gradient expansion.  Because $R^{(1)}$ is constant at first order, equation \eqref{eq:diffglei1o} is a first integral of the well known second-order homogeneous differential equation for $\delta^{(1)}$:
\begin{align}
\delta^{(1)\prime\prime}+\HH \delta^{(1)\prime}-\frac{3}{2}\Omega_{m}\delta^{(1)}=0. \label{eq:secondorderdiff}
\end{align}
The advantage of this fluid-flow approach to relativistic perturbations in the comoving-synchronous gauge is twofold. It is as close as possible to Newtonian perturbation theory (where equation \eqref{eq:secondorderdiff} is exactly the same), with metric perturbations as secondary variables that can be expressed in terms of the density contrast and the curvature and expansion perturbations. Solving equation \eqref{eq:diffglei1o} directly shows that the well known decaying mode $D_{-}$ and the growing mode $D_{+}$ of the solution of \eqref{eq:secondorderdiff} correspond to the homogeneous solution of \eqref{eq:diffglei1o} and the particular solution sourced by the curvature perturbation $R^{(1)}$, respectively:
\begin{align}
&\quad \quad D_{-}+\frac{3}{2}\HH \Omega_{m}D_{-}=0\quad \text{and}\\
&C(\mathbf{x})\left(\HH D_{+}'+\frac{3}{2}\HH^2 \Omega_{m}D_{+}\right)-\frac{1}{4}R^{(1)}=0
\end{align}
with 
\begin{align}
\delta^{(1)}\left(\eta,\mathbf{x}\right)=C_{+}(\mathbf{x})D_{+}(\eta)+C_{-}(\mathbf{x})D_{-}(\eta).\label{eq:d1}
\end{align}

Analogously to equation \eqref{eq:diffglei1o}, we obtain the following for second order combining \eqref{eq:cont} and \eqref{eq:energcon}:
\begin{align}
4\HH \delta^{(2)\prime}+&6\HH^{2}\Omega_{m}\delta^{(2)}- R^{(2)}=2\tht^{(1)2}-2\tht^{(1)i}_{\, j}\tht^{(1)j}_{\, i}-8\HH\delta^{(1)}\tht^{(1)} \label{eq:diffglei2o}.
\end{align}
and \cite{M14}
\begin{align}
R^{(2)\prime}=-4\tht^{(1)i}_{\hspace{3mm}j}R^{(1)j}_{\hspace{3mm}i}=2\left[\partial^i \partial_j \check{\alpha}^{(1)\prime}\partial^j \partial_i \zeta^{(1)}-\nabla^2 \check{\alpha}^{(1)\prime}\nabla^2 \zeta^{(1)}\right]\label{eq:71}
\end{align}
with $g_{ij}=a^2 e^{2\zeta}\left(\delta_{ij}+\check{\alpha}_{,ij}\right)$.

By nature of the perturbative expansion, these second-order equations are sourced by squared first order terms. Given the equivalence of the left-hand side of the systems of equations \eqref{eq:diffglei1o}-\eqref{eq:R1} and \eqref{eq:diffglei2o}-\eqref{eq:71}, it was shown in \cite{M14} that  these equations  are conveniently solved by splitting $\delta^{(2)}$ and $R^{(2)}$ into two parts: 
\begin{equation}
\delta^{(2)}= \delta^{(2)}_h +\delta^{(2)}_p, \quad R^{(2)}= R^{(2)}_h +R^{(2)}_p   \label{eq:72},
\end{equation}
where $\delta_{h}^{(2)}$ and $R_{h}^{(2)}$ are the solutions of the homogeneous parts of \eqref{eq:diffglei2o}-\eqref{eq:71} and $\delta_{p}^{(2)}$ and $R_{p}^{(2)}$ are the particular solutions sourced by the squared first-order terms. In particular, $R_{h}^{(2)}$ is time-independent as $R^{(1)}$ is.

Note that 
 \begin{equation}
  \delta^{(1)} \sim \vartheta^{(1)}  \sim R^{(1)}  \sim \nabla^{2},
  \end{equation}
and this holds true for $\delta^{(2)}_{h}$, $\tht^{(2)}_{h}$, and $R^{(2)}_{h}$, while it is clear from \eqref{eq:diffglei2o}-\eqref{eq:71} that  
\begin{align}
\delta^{(2)}_{p}\sim \tht^{(2)}_{p}\sim R^{(2)}_{p}\sim \nabla^{4}.
\end{align}
Iterating the procedure at higher orders, it follows that at any order $i$,
\begin{equation}
 \delta^{(i)}_{h} \sim \tht^{(i)}_{h}\sim R^{(i)}_{h}\sim \nabla^{2}.
\end{equation}
Therefore, comparing with the equations in the previous section, it should be clear that the leading $\sim \nabla^{2}$ order in the gradient expansion is equivalent to the homogeneous solution of SPT at all orders.

In particular, we can now assume that the Ricci scalar at higher orders can be split into a time-dependent part $R_p ^{(i)}$ and a time-independent part $R_h^{(i)}$. Thus, the gradient expansion offers a unique possibility to compute the homogeneous solution of the evolution equation of $\delta$ at higher orders.
  \section{Third and fourth order}
  \label{sec:34O}
  \subsection{Growing mode solution in the large scale limit}
 Because the evolution equation in the gradient expansion \eqref{eq:diffglei} formally coincides with the evolution equation for the density contrast  \eqref{eq:diffglei1o} at first order, the solution is formally the same. Thus, we solve for the density contrast $\delta$ using the same ansatz \eqref{eq:d1} as we used for the first-order solution. For the growing part of the density contrast sourced by the curvature perturbation we have
\begin{equation}
\delta=D_{+}(\eta)C(\mathbf{x}) \label{eq:delta}.
\end{equation}
 Furthermore, the decaying mode $D_{-}$ is negligible in the matter dominated era. As long as this is well represented by the Einstein-de Sitter model, the growing mode is proportional to the scale factor $a(\eta)$ (see e.g.\ \cite{Bernardeau:2001qr}).

Within the regime of the gradient expansion, the function $C(\mathbf{x})$ is related to the Ricci scalar by \cite{2041-8205-794-1-L11,M14}
\begin{equation}
C(\mathbf{x})=\frac{R}{10 \HH^{2}_{\text{IN}}D_{+\text{IN}}},\label{eq:C}
\end{equation}
where the subscript ``IN'' refers to the evaluation at the time $\eta_{\text{IN}}$ early in the matter-dominated era.  
  \subsection{First, second, third and fourth order solution}
Following the scheme outlined of Section \ref{sec:CPA}, we now compute the homogeneous solution for the second, third, and fourth order density contrast, adding primordial non-Gaussianity to our initial conditions.

We expand $\zeta$ in terms of a Gaussian random field $\zeta^{(1)}$ \cite{Matarrese:2003tk,Lyth:2005fi}:
\begin{equation}
\zeta=\zeta^{(1)}+\frac{3}{5}f_{\text{NL}}\zeta^{(1) 2}+\frac{9}{25}g_{\text{NL}}\zeta^{(1)3}+\frac{27}{125}h_{\text{NL}}\zeta^{(1)4}+\dots.\label{eq:18b}
\end{equation}
where $\f$, $\g$, and $h_{\text{NL}}$ denote the non-Gaussian deviations at different orders.

Now we substitute \eqref{eq:18b} into the series expansion for the Ricci scalar \eqref{eq:RExp}.
For $n=0$ \eqref{eq:RExp} yields the second order expansion of $R$ \cite{M14}. For $n=2$, we obtain $R$ up to the fourth perturbative order: 
\begin{align}
R\simeq& -4\nabla^{2}\zeta +(-2)\left[\left(\nabla \zeta\right)^{2}-4\zeta\nabla^{2}\zeta\right]+\frac{4}{2}\left[2\left(\nabla \zeta\right)^{2}-4\zeta\nabla^{2}\zeta\right]\zeta-\notag\\
&-\frac{4}{3}\left[3\zeta^{2}\left(\nabla\zeta\right)^{2}-4\zeta^{3}\nabla^{2}\zeta\right]+\dots\\
=& -4\nabla^{2}\ze{1}+\left(\nabla\ze{1}\right)^{2}\left[-2-\frac{24}{5}f_{\text{NL}}\right]+\ze{1}\nabla^{2}\ze{1}\left[-\frac{24}{5}f_{\text{NL}}+8\right]+\notag\\
&+\ze{1}\left(\nabla\ze{1}\right)^{2}\left[\frac{216}{25}g_{\text{NL}}+\frac{24}{5}f_{\text{NL}}+4\right]+\zeta^{(1)2}\nabla^{2}\ze{1}\left[-\frac{108}{25}g_{\text{NL}}-8+\frac{48}{5}f_{\text{NL}}\right]+\notag\\
&+\zeta^{(1)2}\left(\nabla\ze{1}\right)^{2}\left[-\frac{1296}{125}h_{\text{NL}}+\frac{324}{125}\g+\frac{72}{25}\f^{2}+\frac{12}{5}\f-4\right]+\notag\\
&+\zeta^{(1)3}\nabla^{2}\ze{1}\left[-\frac{432}{125}h_{\text{NL}}+\frac{288}{25}\g+\frac{144}{25}\f^{2}-\frac{96}{5}\f+\frac{16}{3}\right]+\dots\label{eq:R}
\end{align}
We expand $\delta$ up to fourth order
\begin{equation}
\delta=\delta^{(1)}+\frac{1}{2}\delta^{(2)}+\frac{1}{6}\delta^{(3)}+\frac{1}{24}\delta^{(4)}+\dots
\end{equation}
 and substituting \eqref{eq:R} into \eqref{eq:delta} yields for each order
\begin{align}
\delta^{(1)}_h=&D_{+}(\eta)\frac{1}{10\HH^{2}_{\text{IN}}D_{+\text{IN}}}\left(-4\nabla^{2}\ze{1}\right)\\
\frac{1}{2}\delta^{(2)}_h=&D_{+}(\eta)\frac{1}{10\HH^{2}_{\text{IN}}D_{+\text{IN}}}\notag\\
&\frac{24}{5}\left[-\left(\nabla\ze{1}\right)^{2}\left(\frac{5}{12}+f_{\text{NL}}\right)+\ze{1}\nabla^{2}\ze{1}\left(\frac{5}{3}-f_{\text{NL}}\right)\right]\label{eq:delta2o}\\
\frac{1}{6}\delta^{(3)}_h=&D_{+}(\eta)\frac{1}{10\HH^{2}_{\text{IN}}D_{+\text{IN}}}\notag\\
&\frac{108}{25}\left[\ze{1}\left(\nabla\ze{1}\right)^{2}2\left(g_{\text{NL}}+\frac{5}{9}f_{\text{NL}}+\frac{25}{54}\right)+\zeta^{(1)2}\nabla^{2}\ze{1}\left(-g_{\text{NL}}-\frac{50}{27}+\frac{9}{27}f_{\text{NL}}\right)\right]\label{eq:delta3o}\\
\frac{1}{24}\delta^{(4)}_h=&D_{+}(\eta)\frac{1}{10\HH^{2}_{\text{IN}}D_{+\text{IN}}}\notag\\
&\frac{432}{125}\left[\zeta^{(1)3}\nabla^{2}\ze{1}\left(-h_{\text{NL}}+\frac{10}{3}\g+\frac{5}{3}\f^{2}-\frac{50}{9}\f+\frac{125}{81}\right)+\right.\notag\\
&+\zeta^{(1)2}\left(\nabla\ze{1}\right)^{2}3\left(-h_{\text{NL}}+\frac{1}{4}\g\left.+\frac{5}{18}\f^{2}+\frac{25}{108}\f-\frac{125}{324}\right)\right]\label{eq:delta4o},
\end{align}
where in a general $\Lambda$CDM model $D_{+}$ can be expressed as $D_{+}=\frac{5}{2}\frac{\mathcal{H}^{2}_{\text{IN}}D_{+\,\text{IN}}}{\mathcal{H}^{2}\left(f_{1}\left(\Omega_{m}\right)+\frac{3}{2}\Omega_{m}\right)}$ and $f$ is the standard grow factor:
\begin{align}
f=\frac{D'_{+}}{\HH D_+}.
\end{align}
Eq.\ \eqref{eq:delta2o} is exactly the same solution for the homogeneous part of $\delta$ as in \cite{M14}. The third and fourth order homogeneous solution \eqref{eq:delta3o} are new results. 

Following pioneering work  on second-order perturbations in the nineties \cite{Bruni:1996rg,Matarrese:1997ay,Sonego:1997np,Bruni:1999et}, other second order solutions have been  provided by  \cite{Uggla:2014hva} and \cite{Villa:2015ppa}, cf.\ also \cite{yoo15}, of which the homogeneous part is in accordance with the solution presented here. Solutions up to third order have been derived in \cite{Yoo:2016tcz} using a different gauge, cf.\ also \cite{yoo33,Takahashi:2008yk}.

Furthermore, we are interested in the peaks of the density contrast, thus, we may focus on terms involving $\nabla^{2}\zeta$ as $\nabla\zeta$ vanishes for extremal values.
At second order, the amplitude is decreased by $f_{\text{NL}}^{\text{GR}}=-\frac{5}{3}$ (cf.\ \cite{2041-8205-794-1-L11}), in third order by $g_{\text{NL}}^{\text{GR}}=\frac{50}{27}-\frac{9}{27}f_{\text{NL}}$, and in fourth order by $h_{\text{NL}}^{\text{GR}}=-\frac{10}{9}\g-\frac{5}{9}\f^{2}+\frac{50}{27}\f-\frac{125}{81}$.

It is remarkable that at third order $\f$ contributes to the non-Gaussianity of the density field and at fourth order, additional contributions appear involving  $\f^{2}$ and $\g$.
 The reason becomes quite obvious, when we look at the series expansion of the spatial Ricci scalar $R$ \eqref{eq:RExp}. The third order terms comprise combinations of third and zeroth order, or first and second order. The second order terms contain the non-Gaussianity $\f$ and consequently the combination of first and second order contributes an $\f$ term to the third order result. The fourth order term, on the other hand, comprises combinations of first and third order, second order squared, or two first order and one second order terms, which results in the mixed non-Gaussian terms.
\section{Conclusions}
\label{sec:con}

In this paper we have investigated  non-Gaussian contributions to the density field at very large scales, using a gradient expansion (aka long-wavelength approximation) at  leading order. Our analysis extends the results of \cite{2041-8205-794-1-L11} and \cite{M14} up to fourth order in SPT, within the regime of validity  of the gradient expansion. At second order, our result agrees with the result of \cite{M14}, \cite{Uggla:2014hva}, and \cite{Villa:2015ppa} (cf.\ also \cite{yoo15} for other second order results). At third and fourth order, our solutions are new (other third order results, using a different gauge, were derived by \cite{Yoo:2016tcz}, cf.\ also \cite{yoo33,Takahashi:2008yk}). 

By performing a gradient expansion, we only consider very large scales (of the order of the Hubble radius), at which the spatial gradients are negligible with respect to the time derivatives. We consider spatial gradients up to $\mathcal{O}\left(\nabla^2\right)$, thus including expressions linear in the density contrast, $\delta$, the inhomogeneous expansion, $\tht$, etc.. In this regime, the evolution equations for these variables that characterise inhomogeneity  take the same form as the first-order equations obtained using SPT. In particular, the density contrast $\delta$, as well as the expansion $\vartheta$ are of $\mathcal{O}(\nabla^{2})$, thus any squared term or combination of the two quantities is negligible. At these scales the spatial Ricci scalar remains constant. Using the synchronous-comoving gauge, it is a valid approximation to assume a conformally flat spatial metric, i.e.\ to neglect anisotropic metric perturbations on these scales.
As a consequence, the spatial Ricci scalar can be written as a series expansion, equations \eqref{eq:expR}-\eqref{eq:RExp}.

The evolution equation for the density contrast $\delta$ in the gradient expansion is effectively equivalent to the first-order SPT evolution equation. Therefore, the same ansatz for the growing mode of $\delta$, in which the density contrast is split into a time and a space dependent part, can be used. The spatial amplitude $C(x^{i})$ is proportional to the spatial Ricci scalar, and thereby determined by its nonlinearity. 
Our solution for $\delta$ corresponds to  the homogeneous solution of \cite{M14}, \cite{Uggla:2014hva}, and \cite{Villa:2015ppa}. In the gradient expansion we neglect the terms that source the particular solution.

We show how non-Gaussianity contributes to third and fourth order of SPT within the regime of validity of the gradient expansion. At third order, we obtain terms of order $\mathcal{O}(3)$, $\mathcal{O}(1)\,\mathcal{O}(2)$, and $\mathcal{O}(1)\,\mathcal{O}(1)\mathcal{O}(1)$ in the density contrast. Naturally, the combinations $\mathcal{O}(3)$ and $\mathcal{O}(1)\,\mathcal{O}(2)$  involve terms with $\g$ and $\f$, respectively. Hence, at this order, both $\f$ and $\g$ contribute to the density contrast. At fourth order, we obtain terms containing $h_{\text{NL}}$, $\g$, $\f^2$, and $\f$ in the density contrast.

We should keep in mind that for terms of order $\mathcal{O}(3)$ or higher, our homogeneous solution is not the only contribution in the full solution containing $\f$ terms and other non-Gaussianities. At third order, the particular solution is sourced by terms containing the second order density contrast, which involves $\f$. At fourth order, the source terms for the particular solution will contain the second and third-order density contrast and, thus, terms containing both $\f$ and $\g$.
All these extra terms are however negligible at the large  scales we consider.

In summary, we compute the nonlinear, relativistic contributions in the density field at higher orders at very large scales. In addition we impose initial conditions involving primordial non-Gaussianity up to fourth order. In this context, we see that the nonlinear nature of GR generates both effective non-Gaussian terms and a mixing of the primordial non-Gaussian parameters $\f$, $\g$, and $h_{\text{NL}}$ at higher orders.

Our results should be relevant in the discussion of  higher-order contributions to observables \cite{Bruni:1999et,Yoo:2017svj}, e.g.\ for higher-order statistics such as the bispectrum, cf.\ \cite{Yoo:2016tcz,Takahashi:2008yk}. In addition, they may help in setting initial conditions - and extract relativistic effects - from  simulations of the growth of large scale structure in cosmology, both Newtonian, cf.\ \cite{Bruni:2013mua,Thomas:2015kua} (see also \cite{Wagner:2014aka}) and \cite{Fidler:2017ebh} (and references therein), and in full numerical relativity \cite{PhysRevLett.116.251301,PhysRevLett.116.251302,Macpherson:2016ict,Bentivegna:2016stg,Giblin:2017juu}, cf.\ also  \cite{Adamek:2016zes,Daverio:2016hqi,East:2017qmk}. In turn, fully general relativistic simulations will help to establish the range of validity of higher-order SPT and of the long-wavelength approximation we used in this paper, as well as other nonlinear relativistic approximations such as the post-Friedmann scheme \cite{M15,Rampf:2016wom,Bruni:2013mua}, see also \cite{Thomas:2014aga,Thomas:2015kua,Thomas:2015dfa}, and other approximations \cite{Bruni:1994nf,Hui:1995bw,Bruni:1994nf}. We leave  all of this for future work.
\section*{Acknowledgment} 
The authors are grateful to David Wands, Obinna Umeh  and an anonymous referee for useful comments. H.G. thanks the Faculty of Technology of the University of Portsmouth for support during her PhD studies. M. B. is supported by the UK STFC Grant No. ST/N000668/1.
\newpage
\appendix
\section{Ricci scalar in the gradient expansion}\label{App:A}
In section \ref{sec:GE}, we have assumed that the spatial metric is conformally flat, with $g_{ij}=a^{2}e^{2\zeta}\cg_{ij}\approx a^{2}e^{2\zeta}\delta_{ij}$. On the other hand, a more general metric would read $g_{ij}=a^{2}e^{2\zeta}\cg_{ij}=a^{2}e^{2\zeta}\left(\delta_{ij}+\check{\alpha}_{ij}\right)$ with $\check{\alpha}_{ij}=\partial_{i}\partial_{j}\check{\alpha}$ for scalar perturbations.
It follows that  any contribution from $\check{E}_{ij}$  to the spatial Ricci scalar is of order $\mathcal{O}\left(\nabla^{4}\right)$ and, therefore, it can be neglected in the gradient expansion.  This is easily seen as follows. If two metric are related by a confromal transformation with conformal factor $e^{2\zeta}$, $\gamma_{ij}=e^{2\zeta}\check{\gamma}_{ij}$, then their Ricci scalars are related by \cite{Wald:1984rg}:
\begin{align}
R=e^{-2\zeta}\left[-4\nabla^{2}\zeta-2\left(\nabla\zeta\right)^{2}+\check{R}\right],
\end{align}
where $\check{R}=\check{R}(\cg_{ij})$. $\check{R}$ expressed in terms of the metric $\check{\gamma}_{ij}$ reads
\begin{align}
\check{R}=&\left(\cg^{ij}\cg^{kl}-\cg^{ik}\cg^{jl}\right)\cg_{ij,kl}+\notag\\
&+\cg_{ij,k}\cg_{ab,c}\left(\frac{1}{2}\cg^{ia}\cg^{jc}\cg^{kb}-\frac{3}{4}\cg^{ia}\cg^{jb}\cg^{kc}+\cg^{ia}\cg^{jk}\cg^{bc}+\frac{1}{4}\cg^{ij}\cg^{ab}\cg^{kc}-\cg^{ij}\cg^{ac}\cg^{kb}\right).
\end{align}
At order $\mathcal{O}\left(\nabla^2\right)$, we obtain $\check{R}=0$ given that $\check{\alpha}_{ij}$ is of order $\mathcal{O}\left(\nabla^2 \right)$. The order $\mathcal{O}\left(\nabla^4\right)$ is the first order, at which we obtain non-zero contributions to the spatial Ricci scalar:
\begin{align}
\check{R}\left(\nabla^4\right)=&\left(\delta^{ij}\delta^{kl}-\delta^{ik}\delta^{jl}\right)\check{\alpha}_{ij,kl}+\notag\\
+&\check{\alpha}_{ij,k}\check{\alpha}_{ab,c}\left(\frac{1}{2}\delta^{ia}\delta^{jc}\delta^{kb}-\frac{3}{4}\delta^{ia}\delta^{jb}\delta^{kc}+\delta^{ia}\delta^{jk}\delta^{bc}+\frac{1}{4}\delta^{ij}\delta^{ab}\delta^{kc}-\delta^{ij}\delta^{ac}\delta^{kb}\right).
\end{align}

\section{The curvature perturbation \texorpdfstring{$\zeta$}{z} and the scalar potential \texorpdfstring{$\psi$}{p}}\label{App:B}

We now relate the first-order curvature perturbation $\zeta^{(1)}$ to first order gauge-invariant (GI) Bardeen potentials $\Phi$ and $\Psi$ \cite{PhysRevD.22.1882} and to the Poisson gauge metric variables $\phi_{\rm P}$ and $\psi_{\rm P}$. 

The Ricci scalar $R^{(3)}$ of the comoving slicing, i.e.\ the slicing orthogonal to the irrotational fluid flow with four-velocity $u^{\alpha}$, is a gauge invariant perturbation once we assume a flat FLRW background. We then have (see equation (107) in \cite{Bruni:1992dg})
\begin{align}
R^{(3)}=&-a^{-2}4\nabla^{2}\left(\Psi+\HH V_{S}\right)\label{eq:derzeta1}
\end{align}
with $V_{S}=v+\chi'$ being the GI velocity perturbation.
Again, using a covariant approximation for a perfect fluid with equation of state parameter $w$, one can derive (see equation (127) in \cite{Bruni:1992dg}):
\begin{align}
0=-a\left[3\HH^2 \frac{1}{a^{2}\left(1+w\right)}V_{S}-2a^{-2}\left(\Psi'-\HH \Phi\right)\right]\label{eq:derzeta2}
\end{align}
Combining, \eqref{eq:derzeta1} and \eqref{eq:derzeta2} and using that $\Psi=-\Phi$ yields
\begin{align}
a^2 \delta^{(3)}R=&-4\nabla^{2}\left(\Psi+\HH \frac{2}{3\HH^2\left(1+w\right)}\left(\Psi'-\HH \Phi\right)\right)\\
=&-4\nabla^{2}\left[-\Phi-\frac{2}{3\HH\left(1+w\right)}\left(\Phi'+\HH \Phi\right)\right]\\
=&-4\nabla^{2}\left[-\Phi-\frac{2}{3\left(1+w\right)}\left(\HH^{-1}\Phi'+\Phi\right)\right]\label{eq:deltaR}
\end{align}
We compare \eqref{eq:deltaR} with the first order part of \eqref{eq:expR}:
\begin{align}
-4\nabla^{2}\zeta^{(1)}=&-4\nabla^{2}\left[-\Phi-\frac{2}{3\left(1+w\right)}\left(\HH^{-1}\Phi'+\Phi\right)\right]\\
\zeta^{(1)}=&-\Phi-\frac{2}{3\left(1+w\right)}\left(\HH^{-1}\Phi'+\Phi\right)\label{eq:zetabst}
\end{align}
which coincides with the definition of $\zeta_{BST}$ in \cite{Martin:1997zd}.

In an Einstein-de Sitter universe, we have $\Phi^{\prime}= 0$ and $w=0$. We then obtain for \eqref{eq:zetabst}
\begin{align}
\zeta^{(1)}=-\frac{5}{3}\Phi.\label{eq:zetapsi}
\end{align}
In Poisson gauge, the Bardeen potentials $\Phi$ and $\Psi$ are expressed in terms of the scalar potentials $\RM{\phi}{P}^{(1)}$ and $\RM{\psi}{P}^{(1)}$ as follows:
\begin{align}
\Phi=\RM{\phi}{P}^{(1)}\quad\text{and}\quad \Psi=-\RM{\psi}{P}^{(1)}.\label{eq:phipsi}
\end{align}
For the scalar potentials, the line element reads
\begin{align}
ds^2=a^2\left[-\left(1+2\RM{\phi}{P}^{(1)}\right)d\eta^2+\left(1-2\RM{\psi}{P}^{(1)}\right)\delta_{ij}dx^i dx^j\right].
\end{align}
Therefore, equation \eqref{eq:zetapsi} becomes
\begin{align}
\zeta^{(1)}=-\frac{5}{3}\RM{\psi}{P}^{(1)}.\label{eq:zetapsip}
\end{align}
In the main body of the paper, we have used the synchronous-comoving gauge. A general metric in the synchronous-comoving gauge reads
\begin{align}
ds^2=a^2\left\{-d\eta^2+\left[\left(1-2\RM{\psi}{S}\right)\delta_{ij}+\chi_{\rm{S}ij}\right]dx^i dx^j\right\}\label{eq:metricsynch}
\end{align}
with $\chi_{ij}=\left(\partial_{i}\partial_{j}-\frac{1}{3}\delta_{ij}\nabla^{2}\right)\chi$. At first-order, the metric \eqref{eq:metricsynch} is related to the metric \eqref{eq:metric}, which we used in this paper, via
\begin{align}
e^{2\zeta^{(1)}}\check{\gamma}_{ij}=&\left(1-2\RM{\psi}{S}^{(1)}\right)\delta_{ij}+\chi_{\rm{S}ij}^{(1)}\\
\left(1+2\zeta^{(1)}\right)\delta_{ij}+e^{2\zeta^{(1)}}\check{\alpha}_{,ij}^{(1)}=&\left[1-2\left(\RM{\psi}{S}^{(1)}+\frac{1}{6}\nabla^{2}\RM{\chi}{S}^{(1)}\right)\right]\delta_{ij}+\chi_{\rm{S},ij}^{(1)}\\
2\zeta^{(1)}\delta_{ij}+\check{\alpha}_{,ij}^{(1)}=&-2\left(\RM{\psi}{S}^{(1)}+\frac{1}{6}\nabla^{2}\RM{\chi}{S}^{(1)}\right)\delta_{ij}+\chi_{\rm{S},ij}^{(1)}\\
\Rightarrow \quad \zeta^{(1)}=&-\left(\RM{\psi}{S}^{(1)}+\frac{1}{6}\nabla^{2}\RM{\chi}{S}^{(1)}\right)=-\mathcal{R}_{c}\label{eq:zetapsis}
\end{align}
with $\mathcal{R}_{c}$ being the comoving curvature perturbation and $\check{\alpha}^{(1)}=\RM{\chi}{S}^{(1)}$ at first order. Using the gradient expansion approximation, equation \eqref{eq:zetapsis} becomes
\begin{align}
 \zeta^{(1)}=&-\RM{\psi}{S}^{(1)}.\label{eq:gezetapsis}
\end{align}
To confirm the relation between $\RM{\psi}{P}^{(1)}$ and $\RM{\psi}{P}^{(1)}$ via $\zeta^{(1)}$ from equation \eqref{eq:zetapsip} and \eqref{eq:zetapsis}, we perform a gauge transformation of $\psi^{(1)}$ from Poisson gauge to synchronous-comoving gauge:
\begin{align}
\RM{\phi}{P}^{(1)}=&\RM{\alpha}{PS}^{\prime}+\HH \RM{\alpha}{PS},\\
\RM{\psi}{P}^{(1)}=&\RM{\psi}{S}^{(1)}-\frac{1}{3}\nabla^{2}\RM{\beta}{PS}-\HH \RM{\alpha}{PS}\label{eq:psips}\\
\text{with}\quad &\RM{\alpha}{PS}=\RM{\beta}{PS}^{\prime}=-\frac{1}{2}\RM{\chi}{S}^{(1)\prime}
 \end{align} 
 In \cite{M14}, the first-order scalar potential $\chi$ is expressed in terms of the density contrast $\delta^{(1)}$. For an Einstein-de Sitter universe, we obtain the following relation:
 \begin{align}
 \RM{\chi}{S}^{(1)}=-2\nabla^{-2}\delta^{(1)}_{\rm S}=- \frac{\eta^2}{5}\mathcal{R}_{c}\label{eq:chisy}
 \end{align}
 with $\HH=\frac{2}{\eta}$.
Substituting equation \eqref{eq:chisy} into equation \eqref{eq:psips} yields
\begin{align}
\RM{\psi}{P}^{(1)}=&\RM{\psi}{S}^{(1)}-\frac{1}{3}\nabla^{2}\RM{\beta}{PS}-\HH \RM{\alpha}{PS}\\
=&\RM{\psi}{S}^{(1)}+\frac{1}{6}\nabla^{2}\RM{\chi}{S}^{(1)}+\frac{2}{\eta} \frac{1}{2}\left(-\frac{2\eta}{5}\mathcal{R}_{c}\right)\\
=&\RM{\psi}{S}^{(1)}+\frac{1}{6}\nabla^{2}\RM{\chi}{S}^{(1)}-\frac{2}{5}\mathcal{R}_{c}\\
=&\RM{\psi}{S}^{(1)}+\frac{1}{6}\nabla^{2}\RM{\chi}{S}^{(1)}-\frac{2}{5}\left(\RM{\psi}{S}^{(1)}+\frac{1}{6}\nabla^{2}\RM{\chi}{S}^{(1)}\right)\\
=&\frac{3}{5}\left(\RM{\psi}{S}^{(1)}+\frac{1}{6}\nabla^{2}\RM{\chi}{S}^{(1)}\right)\label{eq:ppp}
\end{align}
Withing the approximation of the gradient expansion, equation \eqref{eq:ppp} simplifies to
\begin{align}
\RM{\psi}{P}^{(1)}=\frac{3}{5}\RM{\psi}{S}^{(1)}\label{eq:psisp},
\end{align}
which is in accordance with equations \eqref{eq:gezetapsis} and \eqref{eq:zetapsip}.
\section{Long and short wavelength split}\label{App:C}
    In the $\Lambda$CDM model, we assume that galaxies evolve in virialised dark matter halos. The halos collapse once the matter density field reaches a critical value. This matter density is determined by the spatial amplitude, $C(\mathbf{x})$, in particular by the nonlinear, spatial Ricci scalar $R$, which comprises of spatial derivatives of $\zeta$. 
While we don't address here issues related to the halo density, we derive formulas for the matter density field, performing a peak-background split, where we decompose $\zeta$ into a longer-wavelength modes $\zl$ and shorter-wavelength modes $\zs$ using $\ze{1}=\zs+\zl$ \cite{2041-8205-794-1-L11}. The short wavelength mode represents modes attributed to local peak formation, whereas the long wavelength modes are assumed to be absorbed into the background. We already did a gradient expansion and by that we are limiting our analysis to large scale wavelengths, $\lambda>\lambda_{\text{min}}$. In the peak-background split, the gradient of the shorter wavelength modes  $\left(\lambda_{\text{min}}<\lambda_{s}<\lambda_{\text{split}}\right)$ still remains small and the gradient of the long wavelength modes ($\lambda_{l}>\lambda_{\text{split}}$) is small enough to be neglected.

Hence, the series expansion of the Ricci scalar \eqref{eq:R} up to fourth order simplifies to
\begin{align}
R\simeq&-4\nabla^{2}\zs-\frac{24}{5}\left(\nabla\zs\right)^{2}\left(\f+\frac{5}{12}\right)-\notag\\
&-\left(\zs+\zl\right)\nabla^{2}\zs\frac{24}{5}\left(\f-\frac{5}{3}\right)-\notag\\
&-\left(\zs+\zl\right)\left(\nabla\zs\right)^{2}\frac{216}{25}\left(\g-\frac{5}{9}\f-\frac{25}{54}\right)-\notag\\
&-\left(\zs+\zl\right)^{2}\nabla^{2}\zs\frac{108}{25}\left(\g-\frac{10}{3}\f+\frac{50}{27}\right)-\notag\\ 
&-\left(\zs+\zl\right)^{2}\left(\nabla\zs\right)^{2}\frac{1296}{125}\left(h_{\text{NL}}-\frac{1}{4}\g-\frac{5}{18}\f^{2}-\frac{25}{108}\f-\frac{125}{324}\right)-\notag\\
&- \left(\zs+\zl\right)^{3}\nabla^{2}\zs\frac{432}{125}\left(h_{\text{NL}}-\frac{10}{3}\g-\frac{5}{3}\f^{2}-\frac{50}{9}\f+\frac{125}{81}\right)\label{eq:Rls}
\end{align}
Substituting this result \eqref{eq:Rls} into the expression for the density contrast \eqref{eq:delta}, where we use \eqref{eq:C} for the spatial function $C(\mathbf{x})$, gives
\begin{align}
\delta=&\frac{1}{\left(\F\right)\HH^{2}}\left[-\nabla^{2}\zs-\frac{6}{5}\left(\nabla\zs\right)^{2}\left(\f+\frac{5}{12}\right)-\right.\notag\\
&-\left(\zs+\zl\right)\nabla^{2}\zs\frac{6}{5}\left(\f-\frac{5}{3}\right)-\notag\\
&-\left(\zs+\zl\right)\left(\nabla\zs\right)^{2}\frac{54}{25}\left(\g-\frac{5}{9}\f-\frac{25}{54}\right)-\notag\\
&-\left(\zs+\zl\right)^{2}\nabla^{2}\zs\frac{27}{25}\left(\g-\frac{10}{3}\f+\frac{50}{27}\right)-\notag\\
&-\left(\zs+\zl\right)^{2}\left(\nabla\zs\right)^{2}\frac{54}{125}\left(h_{\text{NL}}-\frac{1}{4}\g-\frac{5}{18}\f^{2}-\frac{25}{108}\f-\frac{125}{324}\right)-\notag\\
&- \left(\zs+\zl\right)^{3}\nabla^{2}\zs\frac{18}{125}\left(h_{\text{NL}}-\frac{10}{3}\g\left.-\frac{5}{3}\f^{2}-\frac{50}{9}\f+\frac{125}{81}\right)+\dots \right]\label{eq:deltaorder}\\
=&\delta^{(1)}+\frac{1}{2}\delta^{(2)}+\frac{1}{6}\delta^{(3)}+\frac{1}{24}\delta^{(4)}+\dots 
\end{align}
From the above, one can read off the different contributions to the matter density field at different orders.

 It is worth noting that if $\z$ is a Gaussian field,  $\f=\g=h_{\text{NL}}=0$ etc\dots in \eqref{eq:18b},  and therefore $\z=\z^{(1)}$; then the long-wavelength contribution $\zl$ in \eqref{eq:deltaorder}, to the extent that we neglect its gradient,  can be re-absorbed in a coordinate rescaling \cite{2041-8205-794-1-L11,Young:2014ana}. Indeed,
 starting from  \eqref{eq:expR}  we have:
\begin{align}
\label{eq:c4}
R=&e^{-2\z}\left[-4\nabla^{2}\z-2\left(\nabla \z\right)^2\right]\\
\label{eq:c5}
=&e^{-2\ze{1}}\left[-4\nabla^{2}\ze{1}-2\left(\nabla \ze{1}\right)^2\right]\\
\label{eq:c6}
=&e^{-2\left(\zl+\zs\right)}\left[-4\nabla^{2}\zs-2\left(\nabla \zs\right)^2\right]+\mathcal{O}\left(\nabla\zl\right)\\
\approx&e^{-2\zl}R_{s}
\end{align}
with $R_{s}=e^{-2\zs}\left[-4\nabla^{2}\zs-2\left(\nabla \zs\right)^2\right]$. In this case, the forefactor $e^{-2\zl}$ can be absorbed into the background scale factor
\begin{align}
a\rightarrow a_{l}=e^{\zl}a.
\end{align}
For a non-Gaussian $\z$,  it is clear from \eqref{eq:c5}-\eqref{eq:c6} that this coordinate rescaling remains also possible  at second order. At third and higher orders, however, this  is no longer true, because of short-long mixed contributions to the 3-curvature $R$ in \eqref{eq:c4}, terms like $\zl\zs$ arising from the expansion of the  forefactor $e^{-2\z}$   and terms like $\zl\zs\nabla^{2}\zs$ in the square bracket.

\section{Relation between Newtonian and relativistic non-Gaussianities in the matter-dominated era}
\label{App:D}

We now want to relate our relativistic results, obtained with the gradient expansion, with the local-type primordial non-Gaussianity described in a Newtonian fashion, generalising the results in \cite{2041-8205-794-1-L11}. We now focus on the matter-dominated era, assuming therefore $f=\Omega_m =1$ and $\bar{\rho}=\frac{3}{\kappa}\HH^2$.
First, we use the standard expansion for the Newtonian potential \cite{Komatsu:2001rj}:

\begin{equation}
\RM{\phi}{N}=\underbrace{\ph}_{\substack{\phi^{(1)}}}+\underbrace{\fn\left(\ph^{2}-\langle \phi^{2}_{1}\rangle\right)}_{\substack{\frac{1}{2}\phi^{(2)}}}+\underbrace{\g^{\text{N}}\phi_{1}^{3}}_{\substack{\frac{1}{6}\phi^{(3)}}}+\underbrace{h_{\text{NL}}^{\text{N}}\left(\ph^{4}-\langle \ph^{4}\rangle\right)}_{\frac{1}{24}\phi^{(4)}}+\dots
\label{eq:phi}
\end{equation}
Note that $\fn$, $\g^{N}$, and $h_{\text{NL}}^{\text{N}}$ do not refer to primordial non-Gaussianity such as $\f$, $\g$, and $h_{\text{NL}}$, respectively, but to non-Gaussianity in the Newtonian picture at some initial time in the matter dominated era. We now want to split the Newtonian potential into long and short wavelength modes $\phi_1 =\phi_{s}+\phi_{l}$, and substitute them into the Poisson equation. To this end, first consider its gauge-invariant first-order version  in terms of the Bardeen potential $
\Phi$ and the gauge-invariant density perturbation $\RM{\delta}{GI}$ \cite{PhysRevD.22.1882}:
\begin{equation}
\nabla^{2}\Phi=\frac{\kappa}{2}\bar{\rho}\RM{\delta}{GI},
\label{eq:poi1}
\end{equation}
where $\kappa=8\pi G$. Given that $\Phi$ reduces to $\RM{\phi^{(1)}}{P}$ in Poisson gauge and $\RM{\delta}{GI}$ reduces to $\RM{\delta^{(1)}}{S}$ in synchronous-comoving gauge, we get\footnote{See \cite{Wands:2010af} for a discussion of different sign conventions.} \cite{PhysRevD.22.1882,Bruni:1992dg,M14}:
\begin{equation}
\nabla^{2}\RM{\phi}{N}=-\nabla^{2}\RM{\phi^{(1)}}{P}=-\frac{\kappa}{2}\bar{\rho}\RM{\delta^{(1)}}{S},
\label{eq:poi}
\end{equation}
Where the Newtonian potential $\RM{\phi}{N}$ can  been clearly identified with the Poisson gauge metric perturbation $\RM{\phi}{P}$ when  a post-Newtonian-like expansion is used \cite{M15,Rampf:2016wom}. 
As discussed in section \ref{sec:CPA}, the equations in the gradient expansion at leading order formally coincide with those of first-order perturbation theory, while including the homogeneous contributions at all orders. Therefore, we can assume that in this approximation the Poisson equation \eqref{eq:poi} relates $\phi$ and $\delta$ at all orders.
It follows that
\begin{align}
\nabla^{2}\RM{\phi}{N}^{(1)}=&\nabla^{2}\ph=-\frac{\kappa}{2}a^2 \bar{\rho}\delta^{(1)},\\
\frac{1}{2}\nabla^{2}\RM{\phi}{N}^{(2)}=&\nabla^{2}\fn\left(\ph^{2}-\langle\ph^{2}\rangle\right)=-\frac{\kappa}{2}a^2 \bar{\rho}\frac{1}{2}\delta^{(2)},\label{eq:2nd}\\
\frac{1}{6}\nabla^{2}\RM{\phi}{N}^{(3)}=&\g^{\text{N}}\nabla^{2}\ph^{3}=-\frac{\kappa}{2}a^2 \bar{\rho}\frac{1}{6}\delta^{(3)},\;\text{and}\label{eq:3rd}\\
\frac{1}{24}\nabla^{2}\RM{\phi}{N}^{(4)}=&h_{\text{NL}}^{N}\nabla^{2}\left(\ph^{4}-\langle \ph^{4}\rangle \right)=-\frac{\kappa}{2}a^2 \bar{\rho}\frac{1}{24}\delta^{(4)}.\label{eq:4th}
\end{align}
and subsequently, we omit the gradients of the long wavelength terms.
\\
 \noindent \textbf{Second order:}\\
Using the second-order part of equation \eqref{eq:deltaorder} in equation \eqref{eq:2nd} yields 
\begin{align}
&\fn\nabla^{2}\left(\phi_{1}^{2}-\langle \phi_{1}^{2}\rangle\right)=\frac{\kappa}{2}\bar{\rho}\frac{6}{5}\frac{\left(\nabla\zs\right)^{2}\left(\frac{5}{12}+\f\right)+\left(\zs+\zl\right)\nabla^{2}\zs\left(\f-\frac{5}{3}\right)}{\mathcal{H}^{2}\frac{5}{2}}\label{eq:c7}\\
&2\fn\left(\left(\nabla\ps\right)^{2}\hspace{-1mm}+\left(\ps+\pll\right)\nabla^{2}\ps\right)=\frac{18}{25}a^{-2} \left[\left(\nabla\zs\right)^{2}\left(\frac{5}{12}+\f\right)+\left(\zs+\zl\right)\nabla^{2}\zs\left(\f-\frac{5}{3}\right)\right]\label{eq:2omde}
\end{align}
 In the matter-dominated era, the first-order scalar potential $\phi_1$ is linearly related to the first order curvature perturbation $\zeta$ in the following way \cite{2041-8205-794-1-L11}:
\begin{align}
\phi_{1\,i}=\frac{3}{5}\zeta_{i}\quad \text{with the index}\enspace i=l,\,s.\label{eq:phizeta}
\end{align}
In order to discuss our result and compare it to the Newtonian dynamics, we focus on the peaks of the metric perturbations and, therefore, omit the terms involving $\left(\nabla\zs\right)^{2}$ (or $\left(\nabla\ps\right)^{2}$).
\begin{align}
\frac{18}{25}\fn\left(\zs+\zl\right)\nabla^{2}\zs&=\frac{18}{25} \left(\zs+\zl\right)\nabla^{2}\zs\left(\f-\frac{5}{3}\right)\\
\fn&= \left(\f-\frac{5}{3}\right) \label{eq:fnf}
\end{align}
Equation \eqref{eq:fnf} shows that the non-Gaussianity $\fn$ derived in the Newtonian picture consists of the primordial non-Gaussianity $\f$ and an additional term, which has its origin in the nonlinearity of General Relativity. Even if there is no primordial non-Gaussianity $\left(\f=0\right)$, there remains an effective non-Gaussianity of magnitude $\fn=-\frac{5}{3}$. (See also  \cite{2041-8205-794-1-L11,M14})
\\
\textbf{Third order:}\\
Analogously to \eqref{eq:c7}, we combine the third-order part of equation \eqref{eq:deltaorder} with \eqref{eq:3rd} neglecting  any terms involving $\left(\nabla\zs\right)^{2}$:
\begin{align}
3\g^{\text{N}}\left(\ps+\pll\right)^{2}\nabla^{2}\ps=-\frac{\kappa}{2}\bar{\rho}\frac{1}{\frac{5}{2}\HH^{2}}\left[-\frac{27}{25}\left(\zs+\zl\right)^{2}\nabla^{2}\zs\left(\g-\frac{10}{3}\f+\frac{50}{27}\right)\right]
\end{align}
and use the relationship \eqref{eq:phizeta}:
\begin{align}
3\frac{27}{125}\g^{\text{N}}\left(\zs+\zl\right)^{2}\nabla^{2}\zs&-\frac{3}{5} \frac{27}{25}\left(\zs+\zl\right)^{2}\nabla^{2}\zs\left(\g-\frac{10}{3}\f+\frac{50}{27}\right)\\
\g^{\text{N}}&=\left(\g-\frac{10}{3}\f+\frac{50}{27}\right),
\end{align}

which, analogously to the second order approach, is what we aimed to show. We see that we obtain the same non-Gaussian contribution as in \eqref{eq:delta3o}.\\
\\
\textbf{Fourth order:}\\
The recursive process above can be extended to arbitrarily large orders. As an example, here  we use the fourth-order part of \eqref{eq:deltaorder}, substituting it into \eqref{eq:4th} neglecting  any terms involving $\left(\nabla\zs\right)^{2}$:
\begin{align}
4 h_{\text{NL}}^{\text{N}}&\left(\ps+\pll\right)^{3}\nabla^{2}\ps=\notag\\
&=\frac{\kappa}{2}\bar{\rho}\frac{108}{125}\frac{\left(\zs+\zl\right)^{3}\nabla^{2}\zs}{\frac{5}{2}\HH^{2}}\left(h_{\text{NL}}-\frac{10}{3}\g-\frac{5}{3}\f^{2}-\frac{50}{9}\f+\frac{125}{81}\right)
\end{align}
Again we make us of \eqref{eq:phizeta}
\begin{align}
4\left(\frac{3}{5}\right)^4  h_{\text{NL}}^{\text{N}}&\left(\zs+\zl\right)^{3}\nabla^{2}\zs=\frac{3}{5}\frac{108}{125}\left(\zs+\zl\right)^{3}\nabla^{2}\zs\left(h_{\text{NL}}-\frac{10}{3}\g-\frac{5}{3}\f^{2}-\frac{50}{9}\f+\frac{125}{81}\right)\\
 h_{\text{NL}}^{\text{N}}&=h_{\text{NL}}-\frac{10}{3}\g-\frac{5}{3}\f^{2}-\frac{50}{9}\f+\frac{125}{81} 
\end{align}
As in second and third order, we aimed to show that in comparison with the Newtonian gravitational dynamics, we obtain an effective non-Gaussian contribution even with Gaussian primordial initial conditions. $(f_{\text{NL}}=g_{\text{NL}}=h_{\text{NL}}=0)$\\


\bibliographystyle{plain}
\bibliography{MyBib}




\end{document}